\documentstyle[12pt,epsfig]{article}
\def\slash#1{\mbox{$\not \!\! #1$}}

\begin{document}

\pagestyle{empty}
\begin{flushright}
CERN-TH/95-287 \\
ROME1-1122/95\\
\end{flushright}
\vskip 0.5cm
\centerline{\bf ONE-LOOP CORRECTIONS}
\centerline{\bf TO THE TOP, STOP AND GLUINO MASSES IN THE MSSM}
\vskip 0.8cm
\centerline{\bf A. Donini}
\centerline{Theory Division, CERN, 1211 Geneva 23, Switzerland,}
\centerline{Dip. di Fisica,
Universit\`a degli Studi di Roma ``La Sapienza'' and}
\centerline{INFN, Sezione di Roma, P.le A. Moro 2, 00185 Rome, Italy }

\vskip 3cm
\begin{abstract}
We compute the one-loop radiative corrections to the masses of the top quark,
the stop squarks and the gluino in the Minimal Supersymmetric Standard Model.
We include the loops controlled by the strong coupling constant and the top
Yukawa coupling, neglecting those controlled by the $SU(2) \times U(1)$ gauge
couplings and by other Yukawa couplings. We find that the significant
scale-dependence of the renormalization-group-improved tree-level expressions
is almost completely removed. Even for natural choices of the renormalization
scale, corrections can be numerically relevant. Our results should allow more
reliable predictions to be extracted from candidate fundamental theories
beyond the MSSM.
\end{abstract}
\vskip 5cm

\begin{flushleft}
CERN-TH/95-287 \\
November 1995
\end{flushleft}
\vskip 0.5 cm

\newpage
\pagestyle{plain}
\setcounter{page}{1}

\section{Introduction}
\label{sec:intro}

The evaluation of radiative corrections in the Minimal Supersymmetric
Standard Model (MSSM) plays an important role in our understanding of its
phenomenological implications. This is of particular
interest, the MSSM being, at the moment, the most promising extension of the
Standard Model (SM) for the strong and electroweak interactions.

Finite corrections to supersymmetric Ward identities, such as those associated
with Higgs boson masses and couplings \cite{WI}, have a direct impact
on the model predictions. The remaining corrections have more subtle
phenomenological implications, since they can be largely reabsorbed into
redefinition of the many arbitrary parameters of the MSSM. This is no longer
true, however, if one embeds the model into a more fundamental theory (as
superGUTs, supergravities or superstrings) where some or all of the MSSM
parameters can be predicted.

In this paper we compute the one-loop corrections to
the masses of the top quark, the stop squarks and the gluino, going beyond the
leading-log approximation. We present explicit analytical formulae,
keeping in the Feynman rules only the vertices proportional to
the strong coupling constant $g_s$ and to the top Yukawa
coupling $h_t$. This approximation should reproduce
the main contributions to the one-loop radiative corrections.
Comparing our full one-loop results with the renormalization-group-improved
tree-level masses, we find that there can be significant differences,
also in the case of ``natural'' choices for the renormalization scale.
The corrections can be particularly relevant for the lightest stop eigenstate
and for a very light gluino.
In the top quark case, we find that the extra corrections due to loops of
supersymmetric particles can be non-negligible compared to the pure QCD ones.

The paper is organized as follows: in this section, we introduce the notation
by giving the top, stop and gluino tree-level mass formulae; in section
\ref{sec:loop} we introduce the renormalization procedure, we recall the
relation between running and pole masses, and we give the main points of our
calculations; in section \ref{sec:nume} we discuss the numerical results and
their phenomenological implications. The complete results for the
one-loop-corrected top, stop and gluino masses are presented in appendix A.
The explicit solutions of the one-loop renormalization group equations for
all the parameters of relevance to our calculations, in the approximation
where the only non-zero dimensionless couplings are $g_s$ and $h_t$, are
reported in appendix B.

The MSSM \cite{REVIEW} contains three generations of quark
($Q_i,U^c_i,D^c_i$) and lepton ($L_i,E^c_i$) chiral
superfields ($i=1,2,3$), and two Higgs chiral superfields ($H_1,H_2$),
with interactions specified by an $R$-parity-invariant superpotential and a
collection of soft supersymmetry-breaking terms.
As often done in the literature, we shall assume real mass parameters and
neglect all Yukawa couplings apart from the top quark one.
The tree-level expressions for the different masses and couplings
in the MSSM are well known: in the following, we shall use the notation and
conventions of \cite{GUHA1}, including the reversal of the sign of $\mu$ in the
squark sector as specified in the Errata \cite{GUHA1}.

We now recall the tree-level expressions for the top and stop masses
(the tree-level gluino mass is just the explicit soft mass itself).
With no loss of generality, the Vacuum Expectation Values (VEVs) of the Higgs
fields, $v_i = \langle H^0_i \rangle \ (i=1,2)$, can be chosen to be real and
positive. The top quark mass, then, is given by:
\begin{equation}
 m^2_t = h^2_t v^2_2.
\end{equation}
The mass matrix for the stop squarks is:
\begin{equation}
\label{stopmatrix}
M^2_{\tilde t} = \left(
\begin{array}{cc}
m^2_t + \tilde m^2_Q + ( \frac{2}{3} m^2_W - \frac{1}{6} m^2_Z) \cos 2 \beta
& m_t ( A_t - \mu \cot \beta ) \\
m_t ( A_t - \mu \cot \beta ) &
m^2_t + \tilde m^2_T + \frac{2}{3} ( m^2_Z - m^2_W) \cos 2 \beta
\end{array}
\right) ,
\end{equation}
where $\tilde m_Q,\tilde m_T, A_t$ are soft mass terms,
$\tan \beta = v_2/v_1$, $\mu$ is the superpotential mass parameter
and, at the tree level,
\begin{equation}
\label{zetamass}
  m^2_W = \frac{g^2}{2} (v^2_1 + v^2_2), \;\;\;\;\;
  m^2_Z = \frac{g^2 + g^{\prime 2}}{2} (v^2_1 + v^2_2).
\end{equation}
The eigenvalues of (\ref{stopmatrix}) are:
\begin{eqnarray}
\tilde m^2_{t_{1,2}} &=& m^2_t + \frac{1}{2} (\tilde m^2_Q + \tilde m^2_T )
 + \frac{1}{4} m^2_Z \cos 2 \beta  \nonumber \\
&\pm& \sqrt{ \left[ \frac{1}{2} (\tilde m^2_Q - \tilde m^2_T )
 + \frac{1}{12} (8 m^2_W - 5 m^2_Z) \cos 2 \beta \right]^2 + m^2_t ( A_t -
  \mu \cot \beta )^2 } .
\end{eqnarray}
The corresponding eigenstates are:
\begin{equation}
 \tilde t_1 =  \cos \phi_t \tilde t_L + \sin \phi_t \tilde t_R, \;\;\;\;\;
 \tilde t_2 = -\sin \phi_t \tilde t_L + \cos \phi_t \tilde t_R ,
\end{equation}
where the rotation angle $\phi_t$ that diagonalizes the mass matrix is
given by:
\begin{equation}
\tan 2 \phi_t = \frac{m_t (A_t - \mu \cot \beta)}
{[ \frac{1}{2} (\tilde m^2_Q - \tilde m^2_T )
 + \frac{1}{12} (8 m^2_W - 5 m^2_Z) \cos 2 \beta ]} ,
\end{equation}
with $ - \frac{\pi}{2} < \phi_t \le \frac{\pi}{2}$.
Our conventional ordering of the eigenvalues (namely, $\tilde m^2_{t_1} \ge
\tilde m^2_{t_2} $) is obtained with the prescription that $\phi_t$
should have the same sign as the combination $A_t - \mu \cot \beta$.

Summarizing, at the tree level the top, stop and gluino masses are functions
of the following independent input parameters:
$h_t, g, g^{\prime}, m_Z, \tilde m_g, \tilde m_Q, \tilde m_T, A_t, \mu$
 and $\tan \beta$. As we shall see in the next section, other parameters
will appear via the one-loop corrections.

\section{One-loop renormalization}
\label{sec:loop}

At the one-loop level, the pole masses of the top and the gluino are given by:
\begin{eqnarray}
\label{toppolemass}
 m_t &=& \left. [m_t(Q) - \Pi_t(\slash{p},Q) ] \right|_{\slash{p} = m_t},
  \\
\label{gluepolemass}
 \tilde m_g &=& \left. [\tilde m_g(Q) - \Pi_{\tilde g}(\slash{p},Q) ]
     \right|_{\slash{p} = \tilde m_g}.
\end{eqnarray}
In the above equations, $Q$ is the renormalization scale; $m_{t,\tilde g}(Q)$
are the running masses in some renormalization scheme, while
$\Pi_{t,\tilde g}(\slash{p}, Q)$ are the real parts of the renormalized
self-energies in the same scheme. Up to higher-order corrections, the pole
masses do not depend on the scale or on the renormalization scheme.

To evaluate the corrections to the pole masses of the stops, we need to
study the corrected inverse propagator matrix $\hat \Gamma_{ij}$. In the
($\tilde t_1,\tilde t_2$) basis, where $\tilde t_i$ are the tree-level
eigenstates, the expression for the inverse propagator is:
\begin{equation}
\label{inverse}
\hat \Gamma_{ij} = \left(
\begin{array}{cc}
p^2 - \tilde m^2_{t_1}(Q) + \Pi_{11}(p^2,Q) & \Pi_{12}(p^2,Q) \\
  \Pi_{21}(p^2,Q) &  p^2 - \tilde m^2_{t_2}(Q) + \Pi_{22}(p^2,Q)
\end{array}
\right).
\end{equation}
At the one loop, the stop pole masses are:
\begin{eqnarray}
\label{stopmass1}
 \tilde m^2_{t_1}  = \tilde m^2_{t_1}(Q) -
\left. \Pi_{11}(p^2,Q) \right|_{p^2 = \tilde m^2_{t_1}},
  \\
\label{stopmass2}
 \tilde m^2_{t_2}  = \tilde m^2_{t_2}(Q) -
\left. \Pi_{22}(p^2,Q) \right|_{p^2 = \tilde m^2_{t_2}}.
\end{eqnarray}

In eqs. (\ref{toppolemass}), (\ref{gluepolemass}), (\ref{stopmass1}) and
(\ref{stopmass2}), the running masses can be calculated using the tree-level
functional forms (presented in section \ref{sec:intro}), with all the
parameters considered as $Q$-dependent.

\begin{figure}[t]
\vspace{0.1cm}
\centerline{\epsfig{figure=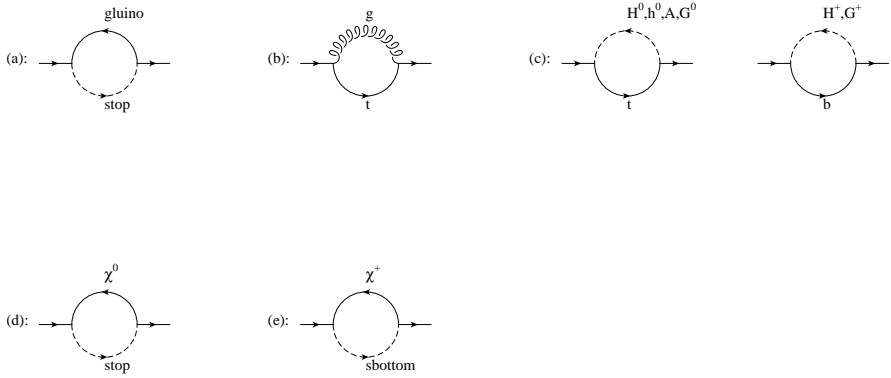,height=5cm,angle=0}}
\caption{\it{One-loop self-energy diagrams for the top:
a) $i \Pi^{\tilde g}_t$; b) $i \Pi^t_t$; c) $i \Pi^H_t$;
d) $i \Pi^{\tilde \chi^0}_t$; e) $i \Pi^{\tilde \chi^+}_t$.
}}
\label{topfeyn}
\end{figure}
\begin{figure}[t]
\vspace{0.1cm}
\centerline{\epsfig{figure=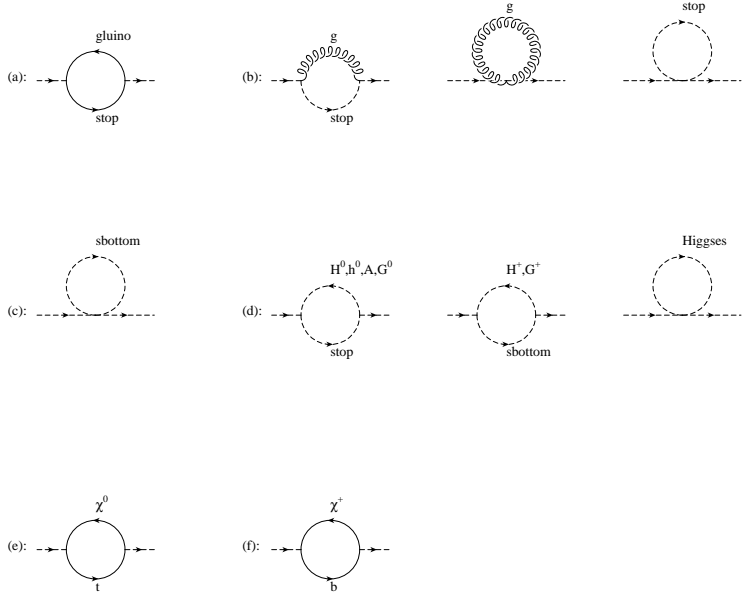,height=8cm,angle=0}}
\caption{\it{
One-loop self-energy diagrams for the stops $(i,j=1,2)$:
a) $i \Pi^{\tilde g}_{ij}$; b) $i \Pi^{\tilde t}_{ij}$;
c) $i \Pi^{\tilde b}_{ij}$; d) $i \Pi^H_{ij}$;
e) $i \Pi^{\tilde \chi^0}_{ij}$; f) $i \Pi^{\tilde \chi^+}_{ij}$.
}}
\label{stopfeyn}
\end{figure}
\begin{figure}[t]
\vspace{0.1cm}
\centerline{\epsfig{figure=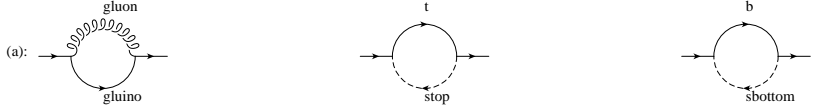,height=1.5cm,angle=0}}
\caption{\it{One-loop self-energy diagrams for the gluino:
a) $i \Pi^{\tilde g}_{\tilde g}$;
b) $i \Pi^{\tilde t}_{\tilde g}$;
c) $i \Pi^{\tilde b}_{\tilde g}$.
}}
\label{gluefeyn}
\end{figure}

To evaluate the one-loop self-energies of top, stops and gluino,
we work in the approximation where all dimensionless couplings other than
$g_s$ and $h_t$ are set to zero, which should take into account the most
relevant one-loop corrections to the studied observables.
The Feynman graphs that represent the self-energies of top, stops and gluino
in this approximation are reported in figs. \ref{topfeyn}, \ref{stopfeyn}
and \ref{gluefeyn}\footnote{The graphs in fig. \ref{gluefeyn}b and
\ref{gluefeyn}c must be counted twice, due to the fact that the gluino
is a Majorana fermion.}. The self-energies for the top ($i \Pi_t$)
and gluino ($i \Pi_{\tilde g}$) read:
\begin{eqnarray}
\label{selftopglue}
i \Pi_t(\slash{p}) &=& i \Pi^{\tilde g}_t +
i \Pi^t_t + i \Pi^H_t + i \Pi^{\tilde \chi^0}_t
+ i \Pi^{\tilde \chi^+}_t,
  \\
i \Pi_{\tilde g}(\slash{p}) &=& i \Pi^{\tilde g}_{\tilde g} +
i \Pi^{\tilde t}_{\tilde g} + i \Pi^{\tilde b}_{\tilde g} \ ,
\end{eqnarray}
while for the stops the self-energies $i \Pi_{ij}$ can be written as:
\begin{eqnarray}
\label{selfstop}
i \Pi_{11}(p^2) &=& i \Pi^{\tilde g}_{11} +
i \Pi^{\tilde t}_{11} + i \Pi^{\tilde b}_{11}
+ i \Pi^H_{11} + i \Pi^{\tilde \chi^0}_{11} + i \Pi^{\tilde \chi^+}_{11},
  \\
i \Pi_{22}(p^2) &=& i \Pi^{\tilde g}_{22} +
i \Pi^{\tilde t}_{22} + i \Pi^{\tilde b}_{22}
+ i \Pi^H_{22} + i \Pi^{\tilde \chi^0}_{22} + i \Pi^{\tilde \chi^+}_{22},
  \\
i \Pi_{12}(p^2) &=& i \Pi^{\tilde g}_{12} +
i \Pi^{\tilde t}_{12} + i \Pi^{\tilde b}_{12}
+ i \Pi^H_{12} + i \Pi^{\tilde \chi^0}_{12} + i \Pi^{\tilde \chi^+}_{12},
  \\
i \Pi_{21}(p^2) &=& i \Pi^{\star}_{12}(p^2).
\end{eqnarray}
They have been grouped in subsets of graphs, to make the results more
transparent and to simplify the numerical study that will be presented in the
next section. The masses of the particles circulating in each subset are
controlled mainly by one input parameter, e.g. the contributions of the graphs
with Higgs and Higgs/squark loops to the stop self-energies (fig.
\ref{stopfeyn}d) can be studied by varying $m^2_A$, while neutralino and
chargino loops (figs. \ref{stopfeyn}e and \ref{stopfeyn}f) depend mainly on
$\mu$.
The self-energies have been calculated both in the Feynman and in the Landau
gauge, with the same results. As usual, in order
to explicitly preserve supersymmetry, we have adopted the $\overline{DR}$
scheme \cite{DRED}. The complete results for the one-loop self-energies of the
top, the two stops and the gluino are reported in appendix A.

It can be seen by looking at the formulae that some new parameters must be
introduced in order to calculate the full one-loop expressions.
The independent parameters that must be added to the ones that appear in the
tree-level relations are $\alpha_s, m_A$ and $\tilde m_B$, where $m_A$ is
the physical mass of the CP-odd neutral Higgs boson and $\tilde m_B$ is the
soft
mass term for the right sbottom squark. The masses of the other Higgs bosons
($H^0, h^0$ and $ H^{\pm}$) and the CP-even mixing angle $\alpha$ can be
deduced with tree-level relations from $m_A, m_Z, m_W$ and $\tan\beta$.
The masses of the sbottoms can be obtained from $\tilde m_Q, \tilde m_B, m_Z,
m_W$ and $\tan\beta$. All these tree-level relations can be found in the
literature.

We have verified that the inclusion of terms proportional to
$g$ and $g^{\prime}$ (namely, $m^2_Z$ and $m^2_W$)
in the tree-level relations for the masses of the particles circulating in
the loops does not significantly modify the results. For this reason, we have
neglected them in the tree-level relations for the charginos and neutralinos
mass matrices. This further approximation greatly simplifies the numerical
implementation of the general formulae reported in appendix A.
The charginos and neutralinos eigenvalues in this approximation are:
\begin{eqnarray}
\label{charginomass}
\tilde  m_{\chi^{+}_i} &=& ( M_2 , \mu) \  \qquad \qquad \qquad
i = 1,2 ,
  \\
\label{neutralinomass}
\tilde   m_{\chi^{0}_j} &=& ( M_1 , M_2 , \mu , - \mu) \ \ \qquad j = 1,4,
\end{eqnarray}
while the rotation matrices are:
\begin{eqnarray}
V_{ij} &=& \left(
\begin{array}{cc}
1 & 0\\
0 & 1
\end{array}
\right) ,
  \\
N_{ij} &=& \left(
\begin{array}{cccc}
1   & 0 & 0                  & 0 \\
0   & 1 & 0                  & 0 \\
0   & 0 & \frac{1}{\sqrt{2}} & - \frac{1}{\sqrt{2}} \\
0   & 0 & \frac{1}{\sqrt{2}} &  \frac{1}{\sqrt{2}}
\end{array}
\right) .
\end{eqnarray}
In this approximation it can be seen, by looking at the formulae reported in
appendix A, that the gaugino soft masses $M_1,M_2$ do not appear explicitly.
Therefore they have not been included in the input parameters set.

In our approximation the electroweak gauge couplings $g$ and $g^{\prime}$
(equivalent to $\sin^2\theta_W$ and $\alpha_{em}$) do not evolve under
the renormalization group equations. They have been fixed at the values
$\sin^2\theta_W = 0.23$ and $\alpha_{em} = \frac{1}{128}$.
Another fixed input parameter is the $Z^0$ physical mass, $m_Z = 91.18$ GeV.
The other two couplings ($\alpha_s,h_t$) have been chosen in order to give,
at the electroweak scale, a value for the top pole mass that lies approximately
in the experimental range. We stress that all the input parameters must be
interpreted as MSSM parameters: e.g. $\alpha_s$ is the MSSM strong coupling
constant, and not the SM one.

\begin{figure}[t]
\vspace{0.1cm}
\centerline{\epsfig{figure=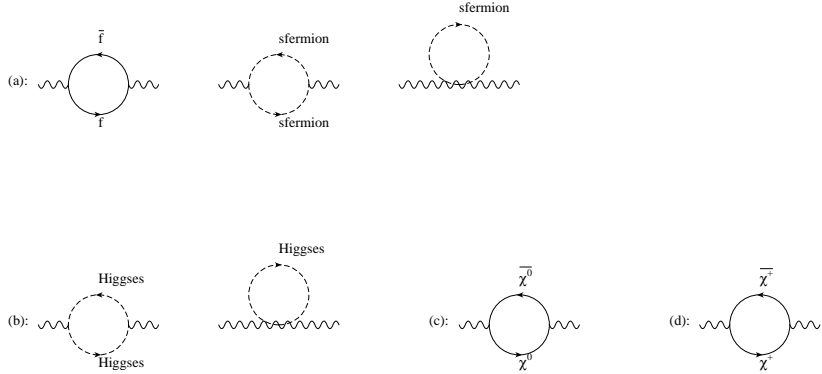,height=5cm,angle=0}}
\caption{\it{One-loop self-energy diagrams for the $Z^0$:
a) $- i g^{\mu\nu} \Pi^f_Z$; b) $-i g^{\mu\nu}\Pi^H_Z$;
c) $-i g^{\mu\nu}\Pi^{\tilde \chi^0}_Z$;
d) $-i g^{\mu\nu}\Pi^{\tilde \chi^+}_Z$.
}}
\label{zetafeyn}
\end{figure}

In order to compare our full one-loop calculations with the results obtained
in the leading-log approximation via the renormalization group equations,
we have used the explicit solution of the RGEs that can be found in their
general form in \cite{KPRZ}. In our approximation we must neglect
terms proportional to dimensionless couplings other than $g_s, h_t$
also in the RGEs, which therefore can be explicitly solved. The details of
the RGEs solutions are reported in appendix B. It must be noted that in
those formulae another two parameters (deducible from the
already discussed set) appear, $v_2$ and $\tilde m_{H_2}$.
For consistency, their expression in terms of the input parameters
has been calculated at the one-loop level. To extract from the input
parameters the one-loop VEV $v_2$, we have calculated the running
$\overline{DR}$ mass $m^2_Z(Q)$ from the physical $Z^0$ mass $m_Z$ using the
following relation:
\begin{equation}
\label{zeta1mass}
m^2_Z(Q) = m^2_Z + \left. \Pi_Z(p^2,Q) \right|_{p^2 = m^2_Z}.
\end{equation}
In the already obtained expression for the one-loop self-energy $\Pi_Z$
\cite{DREHAYA}, we have kept only terms $O(g^2)$
to implement our approximation\footnote{Of course it is not possible to
neglect all the vertices proportional to $g$ and $g^{\prime}$, for this will
result in putting $\Pi_Z = 0$.}. The Feynman graphs that
contribute to the $Z^0$ self-energy in this approximation are reported in
fig. \ref{zetafeyn}.
The decomposition of $\Pi_Z$ on the subsets of graphs is:
\begin{equation}
- i g^{\mu\nu} \Pi_Z(p^2) =
-i g^{\mu\nu} \left[ \Pi^f_Z(p^2) + \Pi^{\tilde \chi^0}_Z(p^2) +
\Pi^{\tilde \chi^+}_Z(p^2) + \Pi^H_Z(p^2) \right].
\end{equation}
The one-loop expression is presented for completeness in appendix A,
jointly with our new one-loop results. The one-loop VEV $v_2$ is then
obtained from $m^2_Z(Q)$ and $\tan \beta$ with the tree-level relation
(\ref{zetamass}). This procedure corresponds to the minimization of the
one-loop effective potential: with this definition of the VEVs, the one-loop
tadpole terms disappear (as the tree-level tadpoles disappear as a
consequence of the tree-level scalar potential minimization). To extract from
the input parameters the one-loop value for $\tilde m_{H_2}$, we have used the
one-loop expressions that relate $\tilde m_{H_1}, \tilde m_{H_2}$ and $m_Z,
m_A$ presented in eqs. (22),(23) and (24) of \cite{BEQZ}.

In the case of the stop squarks we have verified, as a check of the
calculation, that the quadratic divergence of the sum of the full set of
graphs is zero, as it should be in a softly broken supersymmetric theory.
As a further check, we have verified analytically for all the studied masses
that the implicit $Q$-dependence of the running mass cancels with the explicit
$Q$-dependence of the one-loop self-energy. The logarithmic dependence on $Q$
for the renormalized gluino mass has also been checked by a comparison with the
results in \cite{YAMADA}. The one-loop correction to the top quark mass
(including the finite part) in the $\overline{DR}$ scheme for the terms
proportional to $g_s$ has been compared with \cite{BAGGER}. As a consistency
check, we have verified that if the off-diagonal contributions $O(\hbar^2)$ in
eq. (\ref{inverse}) are kept the one-loop-corrected pole masses do not
significantly change.

\section{Numerical analysis}
\label{sec:nume}

In this section we describe in some detail the numerical results obtained for
the one-loop-corrected masses of the top, the two stops and the gluino.

\begin{figure}[t]
\vspace{0.1cm}
\centerline{\epsfig{figure=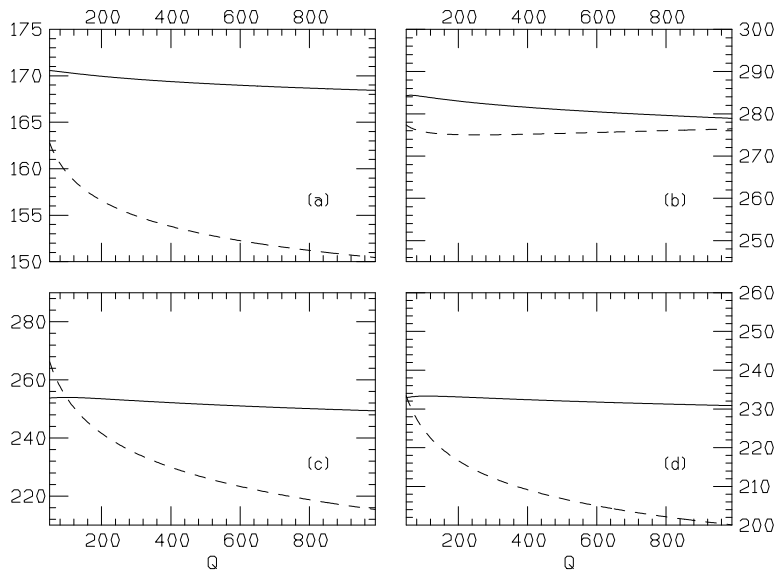,height=11cm,angle=90}}
\caption{\it{Scale-dependence for the one-loop-corrected
masses: a) $t$; b) $\tilde t_1$; c) $\tilde t_2$; d) $\tilde g$.
The solid lines represent the complete one-loop results, while the
dashed lines are the running $\overline{DR}$ masses.
}}
\label{qq_all}
\end{figure}

In fig. \ref{qq_all} we report the scale-dependence of the
$\overline{DR}$ running masses (dashed lines) and of the one-loop-corrected
pole masses (solid lines). Our representative choice of parameters, assigned
conventionally at the scale $Q_0 = 1000$ GeV, is: $\tilde m_Q = \tilde m_T =
\tilde m_B = \tilde m_g = A_t = \mu = m_A = 200$ GeV; $\tan\beta=2$; it can be
seen that, in all four cases, the strong $Q$-dependence of the running mass
$m(Q)$ is significantly reduced. This is an expected feature, due to the
analytic cancellation of the logarithmic dependence in $Q$ between the running
expression and the one-loop self-energy. It must be noted that the residual
scale-dependence present in figs. \ref{qq_all}b and \ref{qq_all}c can be
explained in the following way: using in the numerical calculations the
neutralino and chargino masses reported in eqs. (\ref{charginomass}) and
(\ref{neutralinomass}), we introduce a partial non-cancellation of the
scale-dependence between $\tilde m^2_{t_i} (Q)$ and $\Pi_{ii}(\tilde
m^2_{t_i},Q)$\footnote{By studying the $Q$-dependence of the different sets of
Feynman graphs that appear in fig. \ref{stopfeyn}, we have seen that the
scale-dependence is actually reintroduced adding graphs (e) and (f) to
the others.}. Due to the improved scale-independence, one can check if the
usual procedure adopted to choose the optimal scale to evaluate the
running mass ($Q \simeq m$) can lead to a wrong calculation of the one-loop
effects. This is actually the case, as can be seen for example for
the gluino or the lightest stop mass.

Now we can start a detailed analysis of the complete results, comparing them
with the running masses obtained at the ``natural'' scale $Q \simeq m$.
We find that in many cases the largest corrections to the running
masses come from the loops in which a gluino is circulating.
Therefore we report, for all four masses under consideration,
the dependence of the one-loop-corrected mass on the gluino mass.
We present also the dependence on other parameters in various interesting
cases.
\begin{enumerate}
\item
In fig. \ref{top_all} we report the one-loop-corrected mass for the top.
We have verified that the top mass corrections are almost independent from
$m_A$, which has then been kept fixed at $m_A = 300$ GeV.
The corrections have been calculated at the ``natural'' scale of $Q = 170$ GeV.
We have studied two main situations: the case in which the two stop
states are almost degenerate and the case in which the lightest stop state is
``very'' light (between the experimental lower bounds of $\simeq 50$ GeV and
100 GeV). To obtain degenerate stop states, we have chosen universal soft mass
terms ($\tilde m_Q = \tilde m_T = \tilde m_B = \tilde m_U$) for the squark and
$A_t = \mu \cot\beta= 250$ GeV. This case is illustrated in fig.
\ref{top_all}a, for four representative values of $\tilde m_U$, ($\tilde m_U =
100, 200, 400, 600$ GeV). In the same situation, we have studied the dependence
on $\tan\beta$. The results are reported in fig. \ref{top_all}b, with
$\tilde m_g = 300$ GeV, $\tan\beta=$1--50.
To obtain a light stop state, we have still chosen universal soft mass terms
($\tilde m_Q = \tilde m_T = \tilde m_B = \tilde m_U$) for the squark, but
$A_t = 0$. The dependence of the top mass on $m_A$ in this situation is
reported in fig. \ref{top_all}c, with $m_A=$200--600 GeV. In these last two
figs. $\tilde m_U = 200,400,600$ GeV. It can be seen that in all the cases
presented, the corrections to the running top mass are quite stable and not
very sensitive to the variation of the parameters. The results obtained for
$\tilde m_U = 150$ GeV are reported in fig. \ref{top_all}d. The
corresponding lightest stop mass is approximately $\tilde m_{t_2} < 100$ GeV.
In this last figure, we have studied in detail the single contribution
of the different classes of Feynman graphs to the complete result, in the case
of a ``very light'' $\tilde t_2$. The corrections have been compared with
the QCD correction (represented by the solid line that lies in the middle),
which is the main SM contribution in the approximation $g=g^{\prime}=0$.
The QCD correction to the running mass at $Q=170$ GeV is slightly less than 10
GeV. The dotted lines represent the different contributions added to the QCD
corrections. Starting from the QCD line, we can see the charginos,
neutralinos, Higgses and gluino contributions. It can be seen that the largest
correction (to the QCD result) comes from the gluino loops\footnote{Stop/gluino
loops have already been taken into account in \cite{BAGGER}, with similar
results.}. We stress that, in the different cases examined, the MSSM
corrections can reach half (or more) of the QCD correction. Also, the complete
one-loop results can significantly differ from the running ones.
\begin{figure}[t]   
\centerline{\epsfig{figure=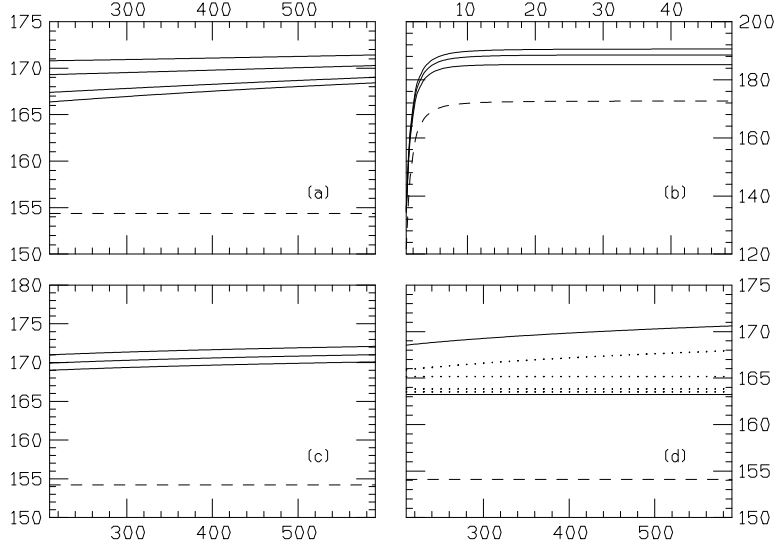,height=11cm,angle=90}}
\caption{\it{One-loop-corrected top mass (in GeV): a) as a function
of the running gluino mass, for $\tilde m_Q = \tilde m_T = \tilde m_B =
100,200,400,600$ GeV from the bottom; $A_t = 250, \mu=500, m_A=300$ GeV;
$\tan\beta=2$. b) as a function of $\tan\beta$, for $\tilde m_Q = \tilde m_T =
\tilde m_B = 200,400,600$ GeV from the bottom; $A_t = 250, \mu=500, m_A=300$;
$\tilde m_g=300$ GeV. c) as a function of $m_A$, for $\tilde m_Q = \tilde m_T =
\tilde m_B = 200,400,600$ GeV from the bottom; $A_t = 0, \mu=500$, $\tilde m_g=
300$ GeV; $\tan\beta=2$. d) as a function of the running gluino mass, for
$\tilde m_Q = \tilde m_T = \tilde m_B = 150$ GeV; $A_t = 0,\mu=500,m_A=300$
GeV; $\tan\beta=2$.
The solid lines are the complete result, the dashed ones represent the running
mass. In (d), the lower solid line is the pure QCD correction; the dotted
lines represent the contributions of (starting from the bottom) charginos,
neutralinos, Higgses and gluino loops added to the pure QCD correction.
}}
\label{top_all}
\end{figure}
\item
In fig. \ref{st1_all} we report the results for the ``heaviest'' stop state
$\tilde t_1$. Also for this particle we have studied the two cases of almost
degenerate stop states and of a ``very light'' $\tilde t_2$. In fig.
\ref{st1_all}a we consider the case of degenerate stop states at three
different values of the universal soft mass term $\tilde m_U = 100,200,400$
GeV. As usual the dashed lines represent the running masses. The three results
have been obtained at the ``natural'' scales $Q=150,240,430$ GeV. In figs.
\ref{st1_all}b and \ref{st1_all}d we have reported on the dependence on $m_A$
and $\mu$ in the case of a ``very light''  $\tilde t_2$ at the scale $Q=300$
GeV (with $\tilde m_U = 170$ GeV, $A_t = 0, \tilde m_g = 300$ GeV). In
particular, in \ref{st1_all}b $\mu=500$ GeV and $m_A=$200--600 GeV while in
\ref{st1_all}d $m_A = 500$ GeV and $\mu=$200--600 GeV. In fig. \ref{st1_all}c,
is reported on the dependence of $\tilde m_{t_1}$ on the gluino mass in the
same situation, with $\tilde m_U=200$ GeV and $Q=320$ GeV. In this case, we
have studied also the contributions coming from the different loops.
The horizontal dotted lines are (from the bottom) the neutralinos, charginos,
Higgses, sbottoms and stops contributions, while the dotted line that
follows the complete result is the gluino contribution. It can be easily seen
that also in this case the most important contribution comes from the gluino
loops. Actually, in all the cases that we have observed, the corrections to
the $\tilde m_{t_1}$ running mass are not very large. We have seen that for
almost degenerate running states, the one-loop-corrected masses can be
interchanged (namely, $\tilde t_2$ can become the heaviest stop state).
\begin{figure}[t]   
\centerline{\epsfig{figure=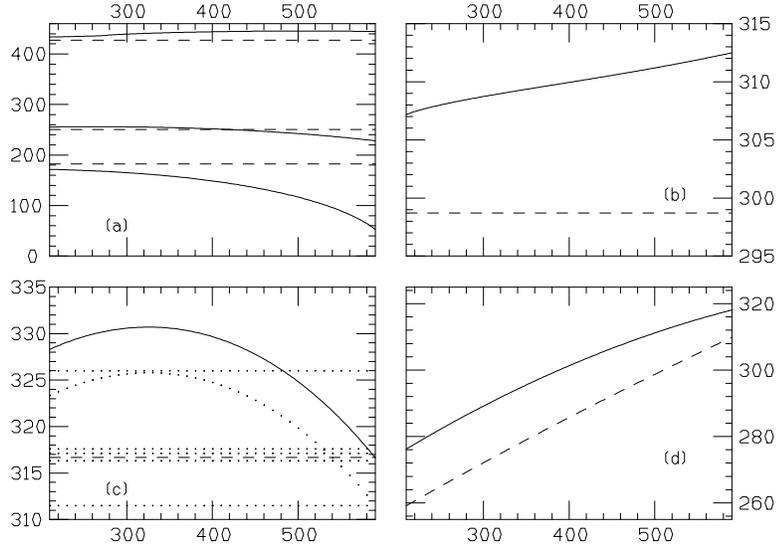,height=11cm,angle=90}}
\caption{\it{One-loop-corrected $\tilde t_1$ mass (in GeV):
a) as a function of the running gluino mass, for $\tilde m_Q = \tilde m_T =
\tilde m_B = 100,200,400$ GeV from the bottom; $A_t = 250, \mu=500, m_A=300$
GeV; $\tan\beta=2$. b) as a function of $m_A$, for $\tilde m_Q = \tilde m_T =
\tilde m_B = 170$ GeV; $A_t = 0, \mu=500$, $\tilde m_g = 300$
GeV; $\tan\beta=2$. c) as a function of the running gluino mass, for
$\tilde m_Q = \tilde m_T = \tilde m_B = 200$ GeV; $A_t = 0, \mu=500, m_A=500$
GeV; $\tan\beta=2$. d) as a function of $\mu$, for $\tilde m_Q = \tilde m_T =
\tilde m_B = 170$ GeV; $A_t = 0, m_A=500$, $\tilde m_g = 300$ GeV;
$\tan\beta=2$.
The solid lines are the complete result, the dashed ones represent the running
mass. In fig. (c) the horizontal dotted lines represent the contributions of
(starting from the bottom) Higgses, sbottoms, charginos, neutralinos and stops
loops. The dotted line with the same shape as the complete result is the
gluino contribution.
}}
\label{st1_all}
\end{figure}
\item
In fig. \ref{st2_all} we report the results for
$\tilde m_{t_2}$ with the same choices of parameters as for $\tilde m_{t_1}$.
In fig. \ref{st2_all}a, it can be seen that $\tilde t_2$ has a behaviour quite
similar to $\tilde t_1$: actually, the one-loop corrections can interchange
the two states ($\tilde t_2$ can become the heaviest one). Also for $\tilde
t_2$, the corrections to the running mass (evaluated at the same ``natural''
scales) are quite small. In figs. \ref{st2_all}b, \ref{st2_all}c and
\ref{st2_all}d we have studied the case of a ``very light'' $\tilde t_2$, at
the ``natural'' scale $Q = 100$ GeV. It can be seen in all three figures
that the one-loop corrections significantly modify the
renormalization-group-improved tree-level results. The most interesting
situation is reported in fig. \ref{st2_all}c. At the ``natural'' scale of
$Q=100$ GeV, we have chosen a universal soft mass term $\tilde m_U= 200$ GeV,
in order to get a sufficiently heavy running mass, $\tilde m_{t_2}(Q) \simeq
150$ GeV. It can be seen from the figure that the one-loop-corrected mass is
significantly different from the running mass, and that can easily fall below
the experimental lower bound of $\simeq 50$ GeV also for a non-super-heavy
gluino mass. This implies that in the case of a light stop state the
running-mass result must be used (also at the ``natural'' scale) with great
care. As usual, we have studied the different contributions to this case.
Starting from the bottom, the horizontal dotted lines are the contributions
coming from neutralinos, charginos, sbottoms, stops and Higgses loops.
The dotted line with the same shape as the complete result represents the
gluino contribution.
\begin{figure}[t]   
\centerline{\epsfig{figure=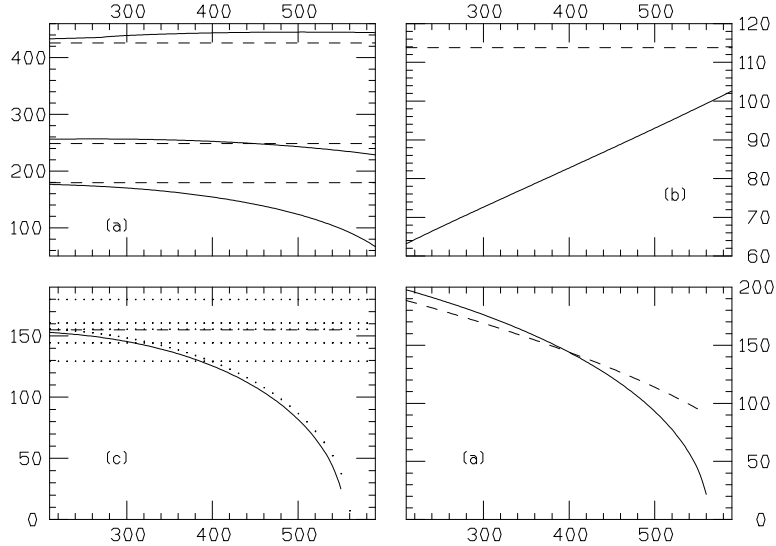,height=11cm,angle=90}}
\caption{\it{One-loop-corrected $\tilde t_2$ mass (in GeV):
a) as a function of the running gluino mass, for $\tilde m_Q = \tilde m_T =
\tilde m_B = 100,200,400$ GeV from the bottom; $A_t = 250, \mu=500, m_A=300$
GeV; $\tan\beta=2$. b) as a function of $m_A$, for $\tilde m_Q = \tilde m_T =
\tilde m_B = 170$ GeV; $A_t = 0, \mu=500$, $\tilde m_g = 300$
GeV; $\tan\beta=2$. c) as a function of the running gluino mass, for
$\tilde m_Q = \tilde m_T = \tilde m_B = 200$ GeV; $A_t = 0, \mu=500, m_A=500$
GeV; $\tan\beta=2$. d) as a function of $\mu$, for $\tilde m_Q = \tilde m_T =
\tilde m_B = 170$ GeV; $A_t = 0, m_A=500$, $\tilde m_g = 300$ GeV;
$\tan\beta=2$.
The solid lines are the complete result, the dashed ones represent the running
mass. In fig. (c), the horizontal dotted lines represent the contributions
of (starting from the bottom) neutralinos, charginos, sbottoms, stops and
Higgses
loops. The dotted line with the same shape as the complete result is the
gluino contribution.
}}
\label{st2_all}
\end{figure}
\item
In fig. \ref{glue_all} we report the results for the one-loop-corrected
gluino mass in two different cases: a heavy gluino and a massless gluino.
Our results agree with the one-loop corrections calculated in \cite{PIEPAP},
and contain as a particular case the results of \cite{FARMAS} for a gluino
massless at tree level. In \ref{glue_all}a and \ref{glue_all}b we have studied
the dependence on $\mu$ and $\tan\beta$ of the corrections to a heavy gluino
mass, for a running mass $\tilde m_g = 300$ GeV at the ``natural'' scale
$Q=300$ GeV. It has been found that the corrections increase with increasing
scalar soft mass terms $\tilde m_Q,\tilde m_T,\tilde m_B$, and with decreasing
$\tan\beta$. They can be of order 10--30\% of the running mass, for stop masses
not exceeding 1 TeV. In figs. \ref{glue_all}c and \ref{glue_all}d we report
results for the massless case (for which we have used $\alpha_s = 0.15$). Here
the situation is reversed, and the corrections increase with decreasing scalar
soft mass terms. It is well known that for a massless gluino the corrections
at one loop are proportional to the mass differences between the squarks. This
has been verified by giving non-universal soft mass terms, keeping the average
scale $\tilde m_U$ at a small value. We have seen that, at the ``natural''
scale $Q=1$ GeV, the corrections to a zero mass gluino can reach 3 or 4 GeV.
Comparing our results with those presented in \cite{FARMAS}, we note that we
have explored a different region of the parameters space\footnote{As long as
we take non-zero gaugino tree-level masses $M_1$ and $M_2$, we can avoid the
experimental bounds induced on $\mu$ by the non-observation at LEP of charginos
lighter than 50 GeV that were considered in \cite{FARMAS}.}.
\begin{figure}[t]   
\centerline{\epsfig{figure=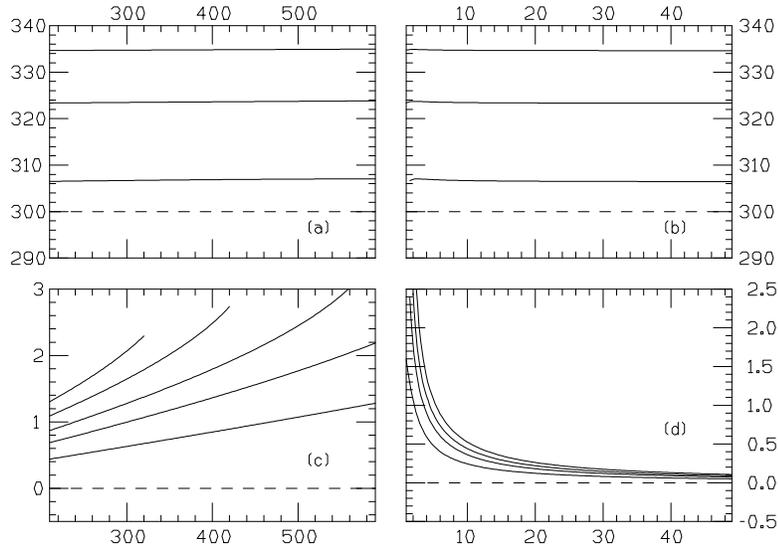,height=11cm,angle=90}}
\caption{\it{One-loop-corrected gluino mass (in GeV):
a) as a function of $\mu$, for $\tilde m_Q = \tilde m_T = \tilde m_B =
200,300,400$ GeV from the bottom; $A_t = 0, m_A=500$ GeV; $\tan\beta=2$.
b) as a function of $\tan\beta$, for $\tilde m_Q = \tilde m_T = \tilde m_B =
200,300,400$ GeV from the bottom; $A_t = 0, m_A = 500, \mu=500$ GeV.
c) as a function of $\mu$, for $\tilde m_Q = \tilde m_T = \tilde m_B =
300,200,150,100,50$ GeV from the bottom; $A_t = 0, m_A=500$ GeV; $\tan\beta=2$.
d) as a function of $\tan\beta$, for $\tilde m_Q = \tilde m_T = \tilde m_B =
300,200,150,100$ GeV from the bottom; $A_t = 0, m_A = 500, \mu=500$ GeV.
The solid lines are the complete result, the dashed ones represent the running
mass.
}}
\label{glue_all}
\end{figure}

\end{enumerate}

\section*{Acknowledgements}
I kindly thank F. Zwirner for continuous help during the completion of this
work and for introducing me to this subject.
I would also like to thank A. Brignole and G. Ridolfi for useful
discussions and suggestions.

\newpage
\section*{Appendix A: One-loop results}
\label{sec:appe2}
In this appendix we report all the explicit results for the one-loop radiative
corrections to the pole masses of the stops, of the gluino and of the top
quark, obtained in the approximation where all the dimensionless couplings
other than $g_s$ and $h_t$ are set to zero.
The corresponding Feynman rules have been deduced from the rules reported in
\cite{REVIEW,FEYN}. It must be noticed that (in the last of \cite{REVIEW})
a factor $\frac{1}{2}$ is missing in the vertices between two neutral
Higgses and two squarks. This factor, due to a combinatorial factor of 2,
cancels if the external legs are the two neutral Higgses giving the Feynman
rule reported in the same reference. The same happens to the
two-gluon/two-squark vertex. In the formulae we have made use of the following
integrals:
\begin{eqnarray*}
  F(1,2,3) &=& {\cal R}e\int^1_0 dz \log \left(\frac{T(1,2,3)}{Q^2} \right)
  \\
  H(1,2,3) &=& {\cal R}e\int^1_0 dz [m^2_1(1-z) + m^2_2z ]
              \log \left(\frac{T(1,2,3)}{Q^2}\right)
  \\
  G(1,2,3) &=& {\cal R}e\int^1_0 dz [m^2_3z(1-z)]
              \log \left(\frac{T(1,2,3)}{Q^2}\right)
  \\
  K(1,2,3) &=& {\cal R}e\int^1_0 dz [m^2_3z(1-z) - m^2_1(1-z) - m^2_2z ]
              \log \left(\frac{T(1,2,3)}{Q^2}\right)
  \\
  L(1,2,3) &=& {\cal R}e\int^1_0 dz [m^2_3 (z-2)^2]
              \log \left(\frac{T(1,2,3)}{Q^2}\right)
  \\
  M(1,2,3) &=& {\cal R}e\int^1_0 dz [m_3 z] \log \left(\frac{T(1,2,3)}{Q^2}
\right)
  \\
  N(1,2,3) &=& {\cal R}e\int^1_0 dz [m_3(1- z)] \log \left(\frac{T(1,2,3)}{Q^2}
\right)
\end{eqnarray*}
where $T(1,2,3) = m^2_1(1-z) + m^2_2 z - m^2_3 z(1-z) - i \epsilon^{\prime}$.
The analytical expressions for these integrals can be found following
\cite{INTE}. They all depend on:
\begin{eqnarray*}
 F(1,2,3) = -2 + \log\left(\frac{m_1 m_2}{Q^2}\right) +
\frac{m^2_1 - m^2_2}{m^2_3}
             log\left(\frac{m_1}{m_2}\right)   \\
+ \frac{1}{m^2_3}
             \sqrt{\mid (m_1 + m_2)^2 - m^2_3\mid\mid (m_1 + m_2)^2 -
             m^2_3\mid} f(1,2,3)
\end{eqnarray*}
where
\begin{equation}
f(1,2,3) = \left\{
\begin{array}{ll}
\log \frac{\sqrt{(m_1 + m_2)^2 -m^2_3}-\sqrt{(m_1 - m_2)^2 -m^2_3}}
{\sqrt{(m_1 + m_2)^2 -m^2_3}+\sqrt{(m_1 - m_2)^2 -m^2_3}}
& {\rm if} \ \ m^2_3 \le (m_1-m_2)^2 \\
  \\
2 \arctan \frac{m^2_3 -(m_1-m_2)^2}{(m_1+m_2)^2 - m^2_3 }
& {\rm if} \ \ (m_1-m_2)^2 < m^2_3 < (m_1+m_2)^2 \\
  \\
\log \frac{\sqrt{m^2_3 - (m_1 - m_2)^2}+\sqrt{m^2_3 - (m_1 + m_2)^2 }}
{\sqrt{m^2_3 - (m_1 - m_2)^2}-\sqrt{m^2_3 - (m_1 + m_2)^2 }}
& {\rm if} \ \ m^2_3 \ge (m_1+m_2)^2
\end{array}
\right.
\end{equation}
The other integrals have the following analytical expressions:
\begin{eqnarray*}
 H(1,2,3) &=& \frac{m^2_1 - m^2_2}{m^2_3}
\left[m^2_1 \left( \log\frac{m^2_1}{Q^2} - 1 \right)
- m^2_2 \left(\log\frac{m^2_2}{Q^2} - 1 \right) \right] \\
&+& \frac{1}{2} \left[m^2_1 + m^2_2 -
\left(\frac{m^2_1 - m^2_2}{m_3}\right)^2\right] F(1,2,3)
  \\
 G(1,2,3) &=& \frac{m^2_3}{18} -\frac{1}{6}
\left[m^2_1 \log\frac{m^2_1}{Q^2}
+ m^2_2 \log\frac{m^2_2}{Q^2} \right] +\frac{2}{3}H(1,2,3) \\
&-& \frac{1}{6} [m^2_1 + m^2_2 - m^2_3]F(1,2,3)
 \\
 K(1,2,3) &=& G(1,2,3) - H(1,2,3)
 \\
 L(1,2,3) &=& m^2_3 \left(4 - \frac{3m^2_1}{m^2_1-m^2_2}\right) F(1,2,3)
            + \frac{3m^2_3}{m^2_1 - m^2_2}H(1,2,3) - G(1,2,3)
 \\
 M(1,2,3) &=& \frac{m_3 m^2_1}{m^2_1 - m^2_2} F(1,2,3)
          - \frac{m_3}{m^2_1-m^2_2} H(1,2,3)
 \\
 N(1,2,3) &=& m_3 F(1,2,3) - M(1,2,3)
\end{eqnarray*}
All these integrals are symmetric under the permutation $1 \rightarrow 2$,
made exception for $L$ and for $M(1,2,3) = N(2,1,3)$.
It has been stressed \cite{INTE} that if some entries of the integral
$F$ are zero, the right expressions are not deducible by just taking the limit
from these formulae. This happens because, if $T(1,2,3)$ has some zero entries,
some of the poles in the logarithm may disappear.
Here we report the main particular cases, for completeness:
\begin{eqnarray*}
 F(1,2,0) &=& -1 + \frac{1}{m^2_1 - m^2_2}
\left[m^2_1 \log \frac{m^2_1}{Q^2}
                 - m^2_2 \log \frac{m^2_2}{Q^2}\right]
 \\
 F(1,1,0) &=& \log\frac{m^2_1}{Q^2} .
\end{eqnarray*}
The particular cases for the other integrals are deducible by taking the limit
for the reported expressions and using the particular expressions for the $F$
integral. This is not true for the $L,M,N$ integrals in the case $m_1 = m_2$:
but also in the pathological case their decomposition over the main
integrals $F,H,G$ is easy to obtain.
Below are reported the expressions for the one-loop radiative corrections
to the pole masses of the stops, the gluino and the top quark.
In these expressions, $N$ is the number of colours ($N=3$),
$(\alpha,\beta) = 1,2,3$ and $(a,b) = 1,\dots,8$.
Moreover, $C_2(R) = \frac{N^2 - 1}{2N}$ and $C_2(G) = N$.
We have used a subscript $c$ for the up-type quarks and squarks different
from top and stop, and a subscript $b$ for all the down-type quarks and
squarks.
In the following, ($\sin\phi_q, \cos\phi_q$) will be abbreviated as
($s_q,c_q$).

\newpage
\begin{enumerate}

\item
Self-energy of the top quark (see fig. \ref{topfeyn}):
\begin{eqnarray*}
i \Pi^t_t(\slash{p}) &= - \frac{i \delta^{\alpha\beta}}{16 \pi^2}
g^2_s C_2(R) & \left[ 2 N(0,t,p) - 4 m_t F(0,t,p) \right] \\
i \Pi^{\tilde g}_t(\slash{p}) &= - \frac{i \delta^{\alpha\beta}}{16 \pi^2}
g^2_s C_2(R) & \left[
M(\tilde g,\tilde t_1,p) - 2\tilde m_g s_t c_t F(\tilde g,\tilde t_1,p)
\right.\\
&&+ \left.  M(\tilde g,\tilde t_2,p) + 2\tilde m_g s_t c_t
F(\tilde g,\tilde t_2,p) \right] \\
i \Pi^H_t(\slash{p}) &= - \frac{i \delta^{\alpha\beta}}{16 \pi^2}
h^2_t & \left\{ \frac{1}{2}\sin^2\alpha \left[ N(H^0,t,p) + m_t F(H^0,t,p))
\right] \right. \\
&&+ \frac{1}{2}\cos^2\alpha \left[N(h^0,t,p) + m_t F(h^0,t,p) \right] \\
&&+ \frac{1}{2}\cos^2\beta  \left[ N(A,t,p) - m_t F(A,t,p)\right] \\
&&+ \frac{1}{2}\sin^2\beta \left[ N(0,t,p) - m_t F(0,t,p)\right] \\
&&+  \left.
\frac{1}{2} \left[ \cos^2\beta M(H^{\pm},b,p)  + \sin^2\beta M(0,b,p) \right]
\right\} \\
i \Pi^{\tilde \chi^0}_t(\slash{p}) &= - \frac{i \delta^{\alpha\beta}}{16 \pi^2}
 h^2_t & \left\{ \frac{1}{2} \sum_j \left[
\mid N_{j4}\mid^2 M(\tilde \chi^0_j, \tilde t_1,p)
+ \tilde m_{\chi^0_j} s_t c_t ( N^2_{j4} + N^{\star^2}_{j4} )
F(\tilde \chi^0_j, \tilde t_1,p) \right] \right.
\\
&&+ \left. \frac{1}{2} \sum_j \left[
\mid N_{j4}\mid^2 M(\tilde \chi^0_j, \tilde t_2,p)
- \tilde m_{\chi^0_j} s_t c_t ( N^2_{j4} + N^{\star^2}_{j4} )
F(\tilde \chi^0_j, \tilde t_2,p) \right] \right\} \\
i \Pi^{\tilde \chi^+}_t(\slash{p}) &= - \frac{i \delta^{\alpha\beta}}{16 \pi^2}
 h^2_t & \left\{ \frac{1}{2} \sum_i  \left[
 c^2_b \mid V_{i2}\mid^2 M(\tilde \chi^+_i, \tilde b_1,p)
+  s^2_b \mid V_{i2}\mid^2 M(\tilde \chi^+_i, \tilde b_2,p) \right] \right\}
\end{eqnarray*}

\item
Self-energies of the stops (see fig. \ref{stopfeyn}):
\begin{enumerate}
\item
\begin{eqnarray*}
i \Pi^{\tilde g}_{11}      &=  \frac{i \delta^{\alpha\beta}}{16 \pi^2}
 & g^2_s C_2(R) \left [ \frac{2}{3} p^2 - 2 (m^2_t + m^2_{\tilde g})
\right. \\
&&- \left. 8 K (\tilde g,t,p) - 4 G (\tilde g,t,p)
- 8 s_t c_t m_t \tilde m_g F (\tilde g,t,p) \right] \\
i \Pi^{\tilde t}_{11}      &=  \frac{i \delta^{\alpha\beta}}{16 \pi^2}
& \left \{  g^2_s C_2(R) \left[ \frac{1}{6} p^2 + \frac{1}{2} \tilde m^2_{t_1}
- 4 s^2_t c^2_t (\tilde m^2_{t_1} - \tilde m^2_{t_2})
-  2 K(0,\tilde t_1,p) \right.\right.\\
 &&+ \left. L(0,\tilde t_1,p) +(1 - 4 s^2_t c^2_t)
 K(\tilde t_1,\tilde t_1,0) + 4 s^2_t c^2_t  K(\tilde t_2,\tilde t_2,0)
\right] \\
&&+ h^2_t \left[ 2 (N+1) s^2_t c^2_t (\tilde m^2_{t_1} - \tilde m^2_{t_2})
           + \tilde m^2_{t_2} \right. \\
&&+ \left.\left. 2 (N+1) s^2_t c^2_t K(\tilde t_1,\tilde t_1,0)
           +(c^4_t + s^4_t - 2 N s^2_t c^2_t) K(\tilde t_2,\tilde t_2,0)
\right] \right \} \\
i \Pi^{\tilde b}_{11}      &=  \frac{i \delta^{\alpha\beta}}{16 \pi^2}
& h^2_t \left\{ s^2_t ( c^2_b \tilde m^2_{b_1} + s^2_b \tilde m^2_{b_2}) +
s^2_t \left[ c^2_b K(\tilde b_1,\tilde b_1,0) + s^2_b
K(\tilde b_2,\tilde b_2,0)\right] \right\} \\
i \Pi^{H}_{11}  &=  \frac{i \delta^{\alpha\beta}}{16 \pi^2} & h^2_t \left[
\frac{1}{2} m^2_{H^0} \sin^2 \alpha + \frac{1}{2} m^2_{h^0} \cos^2 \alpha +
\frac{1}{2} m^2_A \cos^2 \beta + s^2_t m^2_{H^{\pm}} \cos^2 \beta \right.
\\
&&+
\frac{1}{2} \sin^2 \alpha K(H^0,H^0,0) + \frac{1}{2} \cos^2 \alpha K(h^0,h^0,0)
\\
&&+ \frac{1}{2} \cos^2 \beta K(A,A,0) + s^2_t \cos^2 \beta K(H^{\pm},H^{\pm},0)
\\
&&- 2 m^2_t (V^{H^0}_{11})^2 F(H^0,\tilde t_1,p)
   -2 m^2_t (V^{H^0}_{12})^2 F(H^0,\tilde t_2,p)
\\
&&- 2 m^2_t (V^{h^0}_{11})^2 F(h^0,\tilde t_1,p)
   -2 m^2_t (V^{h^0}_{12})^2 F(h^0,\tilde t_2,p)
  \\
&&- \frac{1}{2} \sin^2\beta (V^A_{12})^2 F(A,\tilde t_2,p)
   -\frac{1}{2} \sin^2\beta (V^{G^0}_{12})^2 F(0,\tilde t_2,p)
\\
&&- \sin^2\beta (V^{H^{\pm}}_{11})^2 F(H^{\pm},\tilde b_1,p)
- \sin^2\beta (V^{H^{\pm}}_{12})^2 F(H^{\pm},\tilde b_2,p)
\\
&&- \left.\sin^2\beta (V^{G^{\pm}}_{11})^2 F(0,\tilde b_1,p)
        - \sin^2\beta (V^{G^{\pm}}_{12})^2 F(0,\tilde b_2,p) \right] \\
i \Pi^{\tilde \chi^0}_{11} &=  \frac{i \delta^{\alpha\beta}}{16 \pi^2}
& h^2_t \left[ \frac{1}{3} p^2  - m^2_t - \sum_j \mid N_{j4} \mid^2
\tilde m^2_{\chi^0_j} \right.\\
&&- 4 \sum_j \mid N_{j4}\mid^2 K(\tilde \chi^0_j,t,p)
- 2 \sum_j \mid N_{j4}\mid^2 G(\tilde \chi^0_j,t,p) \\
&& + \left.
2 s_t c_t \sum_j (N^2_{j4} + N^{\star^2}_{j4}) m_t \tilde m_{\chi^0_j}
F(\tilde \chi^0_j,t,p) \right] \\
i \Pi^{\tilde \chi^+}_{11} &=  \frac{i \delta^{\alpha\beta}}{16 \pi^2}
& h^2_t \left[ \frac{1}{3} s^2_t p^2 - s^2_t m^2_b - s^2_t \sum_i
 \mid V_{i2} \mid^2 \tilde m^2_{\chi^+_i}  \right. \\
&&- \left. 4 s^2_t \sum_i \mid V_{i2}\mid^2 K(\tilde \chi^+_i,b,p)
  - 2 s^2_t \sum_i \mid V_{i2}\mid^2 G(\tilde \chi^+_i,b,p) \right]
\end{eqnarray*}
\item
\begin{eqnarray*}
i \Pi^{\tilde g}_{22}      &=  \frac{i \delta^{\alpha\beta}}{16 \pi^2}
& g^2_s C_2(R) \left [ \frac{2}{3} p^2 - 2(m^2_t + m^2_{\tilde g})
\right. \\
&&- \left. 8 K (\tilde g,t,p) - 4 G (\tilde g,t,p) + 8 s_t c_t m_t \tilde m_g
F (\tilde g,t,p) \right] \\
i \Pi^{\tilde t}_{22}      &=  \frac{i \delta^{\alpha\beta}}{16 \pi^2}
& \left \{ g^2_s C_2(R) \left[ \frac{1}{6} p^2 + \frac{1}{2} \tilde m^2_{t_2}
+ 4 s^2_t c^2_t (\tilde m^2_{t_1} - \tilde m^2_{t_2})
-  2 K(0,\tilde t_2,p) \right.\right.\\
&&+ \left. L(0,\tilde t_2,p) +(1 - 4 s^2_t c^2_t)
 K(\tilde t_2,\tilde t_2,0) + 4 s^2_t c^2_t  K(\tilde t_1,\tilde t_1,0)
\right] \\
&&+ h^2_t \left[ - 2 (N+1) s^2_t c^2_t (\tilde m^2_{t_1} - \tilde m^2_{t_2})
           + \tilde m^2_{t_1} \right. \\
&&+ \left.\left. 2 (N+1) s^2_t c^2_t K(\tilde t_2,\tilde t_2,0)
           +(c^4_t + s^4_t - 2 N s^2_t c^2_t) K(\tilde t_1,\tilde t_1,0)
\right] \right \} \\
i \Pi^{\tilde b}_{22}      &=  \frac{i \delta^{\alpha\beta}}{16 \pi^2}
& h^2_t \left\{ c^2_t ( c^2_b \tilde m^2_{b_1} + s^2_b \tilde m^2_{b_2}) +
c^2_t \left[ c^2_b K(\tilde b_1,\tilde b_1,0) + s^2_b
K(\tilde b_2,\tilde b_2,0)\right] \right\} \\
i \Pi^{H}_{22}  &=  \frac{i \delta^{\alpha\beta}}{16 \pi^2} & h^2_t \left[
\frac{1}{2} m^2_{H^0} \sin^2 \alpha + \frac{1}{2} m^2_{h^0} \cos^2 \alpha +
\frac{1}{2} m^2_A \cos^2 \beta + c^2_t m^2_{H^{\pm}} \cos^2 \beta \right.
\\
&&+
\frac{1}{2} \sin^2 \alpha K(H^0,H^0,0) + \frac{1}{2} \cos^2 \alpha K(h^0,h^0,0)
\\
&&+ \frac{1}{2} \cos^2 \beta K(A,A,0) + c^2_t \cos^2 \beta K(H^{\pm},H^{\pm},0)
\\
&&- 2 m^2_t (V^{H^0}_{21})^2 F(H^0,\tilde t_1,p)
   -2 m^2_t (V^{H^0}_{22})^2 F(H^0,\tilde t_2,p)
\\
&&- 2 m^2_t (V^{h^0}_{21})^2 F(h^0,\tilde t_1,p)
   -2 m^2_t (V^{h^0}_{22})^2 F(h^0,\tilde t_2,p)
  \\
&&- \frac{1}{2} \sin^2\beta (V^A_{12})^2 F(A,\tilde t_1,p)
   -\frac{1}{2} \sin^2\beta (V^{G^0}_{12})^2 F(0,\tilde t_1,p)
\\
&&- \sin^2\beta (V^{H^{\pm}}_{21})^2 F(H^{\pm},\tilde b_1,p)
  - \sin^2\beta (V^{H^{\pm}}_{22})^2 F(H^{\pm},\tilde b_2,p)
\\
&&- \left.\sin^2\beta (V^{G^{\pm}}_{21})^2 F(0,\tilde b_1,p)
        - \sin^2\beta (V^{G^{\pm}}_{22})^2 F(0,\tilde b_2,p)
\right] \\
i \Pi^{\tilde \chi^0}_{22} &=  \frac{i \delta^{\alpha\beta}}{16 \pi^2}
&h^2_t \left[ \frac{1}{3} p^2  - m^2_t - \sum_j \mid N_{j4} \mid^2
\tilde m^2_{\chi^0_j} \right.\\
&&- 4 \sum_j \mid N_{j4}\mid^2 K(\tilde \chi^0_j,t,p)
- 2 \sum_j \mid N_{j4}\mid^2 G(\tilde \chi^0_j,t,p) \\
&&- \left.
 2 s_t c_t \sum_j (N^2_{j4} + N^{\star^2}_{j4}) m_t \tilde m_{\chi^0_j}
F(\tilde \chi^0_j,t,p) \right] \\
i \Pi^{\tilde \chi^+}_{22} &=  \frac{i \delta^{\alpha\beta}}{16 \pi^2}
&h^2_t \left[ \frac{1}{3} c^2_t p^2 - c^2_t m^2_b - c^2_t \sum_i
 \mid V_{i2} \mid^2 \tilde m^2_{\chi^+_i}  \right. \\
&&- \left. 4 c^2_t \sum_i \mid V_{i2}\mid^2 K(\tilde \chi^+_i,b,p)
  - 2 c^2_t \sum_i \mid V_{i2}\mid^2 G(\tilde \chi^+_i,b,p) \right]
\end{eqnarray*}
\item
\begin{eqnarray*}
i \Pi^{\tilde g}_{12} &=  \frac{i \delta^{\alpha\beta}}{16 \pi^2}
& g^2_s C_2(R) \cos 2 \phi_t \left[  - 4 m_t \tilde m_g F(\tilde m_g,t,p)
\right] \\
i \Pi^{\tilde t}_{12} &=  \frac{i \delta^{\alpha\beta}}{16 \pi^2}
& \left\{  g^2_s C_2(R) \cos 2 \phi_t \left[
 -  s_t c_t ( \tilde m^2_{t_1} - \tilde m^2_{t_2} ) \right.\right. \\
&&- \left. 2 s_t c_t
( K(\tilde t_1, \tilde t_1, 0 ) - K(\tilde t_2, \tilde t_2, 0 ) )\right] \\
&&+ h^2_t \left[
(N+1)s_t c_t \cos 2 \phi_t ( \tilde m^2_{t_1} - \tilde m^2_{t_2} ) \right.\\
&&+ \left. \left. (N+1) s_t c_t \cos 2\phi_t ( K(\tilde t_1,\tilde t_1,0) -
K(\tilde t_2,\tilde t_2,0)) \right] \right\} \\
i \Pi^{\tilde b}_{12} &=  \frac{i \delta^{\alpha\beta}}{16 \pi^2}
& h^2_t \left\{ s_t c_t (c^2_b \tilde m^2_{b_1} + s^2_b \tilde m^2_{b_2} )
\right. \\
&&+ \left. s_t c_t \left[ c^2_b K(\tilde b_1,\tilde b_1,0)
+ s^2_b K(\tilde b_2,\tilde b_2,0)\right] \right\} \\
i \Pi^H_{12} &=  \frac{i \delta^{\alpha\beta}}{16 \pi^2}
& h^2_t \left[ m^2_{H^{\pm}} \cos^2 \beta s_t c_t
+ \cos^2 \beta s_t c_t K(H^{\pm},H^{\pm},0) \right. \\
&&- 2 m^2_t (V^{H^0}_{11}V^{H^0}_{12}) F(H^0,\tilde t_1,p)
   -2 m^2_t (V^{H^0}_{12}V^{H^0}_{22}) F(H^0,\tilde t_2,p)
\\
&&- 2 m^2_t (V^{h^0}_{11}V^{h^0}_{12}) F(h^0,\tilde t_1,p)
   -2 m^2_t (V^{h^0}_{12}V^{h^0}_{22}) F(h^0,\tilde t_2,p)
  \\
&&- \sin^2 \beta (V^{H^{\pm}}_{11}V^{H^{\pm}}_{12}) F(H^{\pm},\tilde b_1,p)
  - \sin^2 \beta (V^{H^{\pm}}_{12}V^{H^{\pm}}_{22}) F(H^{\pm},\tilde b_2,p)
\\
&&- \left.
  \sin^2 \beta (V^{G^{\pm}}_{11}V^{G^{\pm}}_{12}) F(0,\tilde b_1,p)
- \sin^2 \beta (V^{G^{\pm}}_{12}V^{G^{\pm}}_{22}) F(0,\tilde b_2,p)
\right] \\
i \Pi^{\tilde \chi^0}_{12} &=  \frac{i \delta^{\alpha\beta}}{16 \pi^2}
&h^2_t \left[ 2 \sum_j (c^2_t N^{\star^2}_{j4} - s^2_t N^2_{j4} ) m_t
\tilde m_{\chi^0_j} F(\tilde \chi^0_j,t,p) \right] \\
i \Pi^{\tilde \chi^+}_{12} &=  \frac{i \delta^{\alpha\beta}}{16 \pi^2}
& h^2_t \left[ \frac{2}{3}s_t c_t p^2 - 2 s_t c_t m^2_b - 2 s_t c_t \sum_i
 \mid V_{i2} \mid^2 \tilde m^2_{\chi^+_i}  \right. \\
&&-  \left. 2 s_t c_t \sum_i \mid V_{i2}\mid^2 (K(\tilde \chi^+_i,b,p)
+ G(\tilde \chi^+_i,b,p)) \right] \\
\end{eqnarray*}
\end{enumerate}
In these expressions, we have used the following definitions for the
Higgs--squark--squark vertices in the ($\tilde t_1, \tilde t_2$) basis:
\begin{eqnarray*}
 V^{H^0}_{11} &=& \sin \alpha
           + \frac{1}{m_t} s_t c_t ( A_t \sin \alpha - \mu \cos \alpha)
  \\
 V^{H^0}_{12} &=& V^{H^0}_{21} =
  \frac{1}{2m_t} \cos 2 \phi_t ( A_t \sin \alpha - \mu \cos \alpha)
  \\
 V^{H^0}_{22} &=& \sin \alpha
           - \frac{1}{m_t} s_t c_t ( A_t \sin \alpha - \mu \cos \alpha)
  \\
 V^A_{12} &=& V^A_{21} = ( A_t \cot \beta + \mu )
  \\
 V^{G^0}_{12} &=& V^{G^0}_{21} = ( A_t  - \mu \cot \beta)
  \\
 V^{H^{\pm}}_{11} &=&
m_t c_t c_b \cot \beta + s_t c_b (A_t \cot\beta + \mu)
  \\
 V^{H^{\pm}}_{12} &=&
-m_t c_t s_b \cot \beta - s_t s_b (A_t \cot\beta + \mu)
  \\
 V^{H^{\pm}}_{21} &=&
-m_t s_t c_b \cot \beta + c_t c_b (A_t \cot\beta + \mu)
  \\
 V^{H^{\pm}}_{22} &=&
m_t s_t s_b \cot \beta - c_t s_b (A_t \cot\beta + \mu)
  \\
 V^{G^{\pm}}_{11} &=&
m_t c_t c_b + s_t c_b (A_t - \mu \cot\beta)
  \\
 V^{G^{\pm}}_{12} &=&
-m_t c_t s_b - s_t s_b (A_t - \mu \cot\beta)
  \\
 V^{G^{\pm}}_{21} &=&
-m_t s_t c_b + c_t c_b (A_t - \mu \cot\beta)
  \\
 V^{G^{\pm}}_{22} &=&
m_t s_t s_b  - c_t s_b (A_t - \mu \cot\beta)
\end{eqnarray*}
The vertices $V^{h^0}_{ij}$ can be obtained by $V^{H^0}_{ij}$,
replacing $\sin \alpha \rightarrow \cos \alpha$ and
$\cos \alpha \rightarrow - \sin \alpha$.

\item
Self-energy of the gluino (see fig. \ref{gluefeyn}):
\begin{eqnarray*}
i \Pi^{\tilde g}_{\tilde g} (\slash{p}) &=&
-\frac{i g^2_s \delta^{ab} }{16 \pi^2}
\left[ 2 C_2(G) N(0,\tilde g,p) - 4 \tilde m_g C_2(G) F(0,\tilde g,p) \right]
\\
i \Pi^{\tilde t}_{\tilde g} (\slash{p}) &=&
-\frac{i g^2_s \delta^{ab} }{16 \pi^2}
\sum^3_{i=1} \left[ N(\tilde t_{1_i},t_i,p)
- 2 m_{t_i} s_{t_i} c_{t_i} F(\tilde t_{1_i},t_i,p)
+ N(\tilde t_{2_i},t_i,p) + 2 m_{t_i} s_{t_i} c_{t_i}  F(\tilde t_{2_i},t_i,p)
\right] \\
i \Pi^{\tilde b}_{\tilde g} (\slash{p}) &=&
-\frac{i g^2_s \delta^{ab} }{16 \pi^2}
\sum^3_{i=1} \left[ N(\tilde b_{1_i},b_i,p)
- 2 m_{b_i} s_{b_i} c_{b_i} F(\tilde b_{1_i},b_i,p)
+ N(\tilde b_{2_i},b_i,p) + 2 m_{b_i} s_{b_i} c_{b_i}  F(\tilde b_{2_i},b_i,p)
\right]
\end{eqnarray*}
In this expression, $i$ is the generation index,
and $t,b$ stand for top-type and bottom-type.

\item
Self-energy of the $Z^0$ vector boson (see fig. \ref{zetafeyn}):
\begin{eqnarray*}
\frac{\Pi^f_Z(p^2)}{K m^2_Z} &=& \sum^3_{i=1}
 \left\{ \frac{1}{2} m^2_{f_i}  F(f_i,f_i,Z)
+ 2( c^2_{f_i} c_{f_i L} - s^2_{f_i} c_{f_i R})^2 \tilde m^2_{f_{i_1}}
F(\tilde f_{i_1},\tilde f_{i_1},Z) \right.  \\
&&+ 2( s^2_{f_i} c_{f_i L} - c^2_{f_i} c_{f_i R})^2 \tilde m^2_{f_{i_2}}
F(\tilde f_{i_2},\tilde f_{i_2},Z)
+ 4 s^2_{f_i} c^2_{f_i} (c_{f_i L} + c_{f_i R})^2
H(\tilde f_{i_1},\tilde f_{i_2},Z) \\
&&- \left. 2(c^2_{f_i} c^2_{f_i L} + s^2_{f_i} c^2_{f_i R})
\tilde m^2_{f_{i_1}} F(\tilde f_{i_1},\tilde f_{i_1},0)
-2(s^2_{f_i} c^2_{f_i L} + c^2_{f_i} c^2_{f_i R}) \tilde m^2_{f_{i_2}}
F(\tilde f_{i_2},\tilde f_{i_2},0) \right\} \\
\frac{\Pi^{\tilde \chi^0}_Z(p^2)}{K m^2_Z} &\simeq& 0 \\
\frac{\Pi^{\tilde \chi^+}_Z(p^2)}{K m^2_Z} &\simeq& 0 \\
\frac{\Pi^H_Z(p^2)}{K m^2_Z} &=& \frac{1}{3}  \left[
\frac{1}{2} H(H^0,A,Z) + \frac{1}{2} \cos^2(2 \theta_W) H (H^{\pm},H^{\pm},Z)
\right.\\
&&- \left. \frac{1}{4} H(H^0,H^0,0) - \frac{1}{4}H(A,A,0)
- \frac{1}{2} \cos^2(2 \theta_W) H(H^{\pm},H^{\pm},0) \right]
\end{eqnarray*}
In these expressions we have used:
\begin{eqnarray*}
K &=& \frac{3}{16 \pi^2} \frac{g^2}{m^2_W}
 \\
c_{fL} &=& T_{3f} - e_f \sin^2\theta_W
  \\
c_{fR} &=& e_f \sin^2\theta_W
\end{eqnarray*}
where $f$ represents top- and bottom-type quarks, neutrinos and charged
leptons (with $e_t = \frac{2}{3},e_b = -\frac{1}{3}, e_{\nu} = 0,e_l = -1$
and $T_{3t}=T_{3\nu}=\frac{1}{2},T_{3b}=T_{3l}=-\frac{1}{2}$).
This result has already been obtained \cite{DREHAYA}.
We have reported it here for completeness, in the approximated form
in which terms proportional to $O(m^4_Z)$ are neglected.
\end{enumerate}

\section*{Appendix B: Renormalization group evolution}
\label{sec:appe1}
In this appendix we report the explicit solutions of the
one-loop renormalization group equations for the parameters relevant to our
calculations, in the approximation where the only non-vanishing dimensionless
couplings are $g_s$ and $h_t$. In the following, $t =\log(\frac{Q^2_0}{Q^2})$,
where $Q_0$ is an arbitrary reference scale. The RGEs of interest can be
extracted from the general formulae of \cite{IBAN}.
It must be stressed that, in accordance with our sign and phase convention
\cite{GUHA1}, the correct equations that must be solved
correspond to the equations reported in the
second paper of \cite{IBAN} with the substitution $A_q \rightarrow -A_q$.
Following the notation of \cite{KPRZ}, we introduce a set of parameters:
\begin{eqnarray}
 Z_3 &=& \left( 1 + \frac{b_3}{4 \pi} \alpha_s(0) t \right)^{-1}
  \\
 E &=& Z^{\frac{16}{9}}_3
  \\
 F &=& \int^t_0 dt^{\prime} E(t^{\prime}) = -
\frac{3}{7}\frac{4 \pi}{\alpha_s(0)} \left( 1 - \frac{E}{Z_3} \right)
\end{eqnarray}
where $b_3 = -3$.
It can be seen that, in our approximation, the integral $F$ can be easily
solved. Introducing the variable
\begin{equation}
x(t) = \frac{3}{2 \pi} \frac{F}{E} \alpha_t(t)
\end{equation}
the explicit solutions for a first set of equations are:
\begin{eqnarray}
 \alpha_s(t) &=& \alpha_s(0) Z_3
  \\
 \alpha_t(t) &=& \frac{\alpha_t(0) E}{1 + \frac{3}{2 \pi} \alpha_t(0) F}
  \\
 \mu^2(t) &=& \mu^2(0) \left(\frac{\alpha_t(t)}{\alpha_t(0)}
             \right)^{\frac{1}{2}} Z^{-\frac{8}{9}}_3
  \\
 v^2_2(t) &=& v^2_2(0) \left(\frac{\alpha_t(t)}{\alpha_t(0)}\right)^{
            -\frac{1}{2}} Z^{\frac{8}{9}}_3
  \\
 A_t(t) &=& A_t(0) (1 - x) - \tilde m_g(0)
\left[ \frac{16}{3}\frac{\alpha_s(0)}{4 \pi} t Z_3 - \frac{1}{2}
 2(t \frac{E}{F} - 1) x \right]
  \\
 \tilde m_g(t) &=& \tilde m_g(0) Z_3 \ .
\end{eqnarray}
The explicit solutions for the soft masses RGEs depend on the quantity:
\begin{eqnarray}
\Delta m^2_t &=& \frac{1}{2} ( \tilde m^2_Q(0) + \tilde m^2_T(0) +
 \tilde m^2_{H_2}(0) ) x  \nonumber \\
&-& \frac{1}{2}A_t(0) x (1-x) \left[ 2 \left(t \frac{E}{F} - 1 \right)
\tilde m_g(0) - A_t(0) \right] \nonumber \\
&+& \frac{1}{2} \tilde m^2_g(0)x \left[
t \frac{E}{F} \frac{16}{3}\frac{\alpha_s(0)}{4 \pi} t Z_3
- \left(t \frac{E}{F} - 1 \right)^2 x \right] \ .
\end{eqnarray}
Then, we obtain:
\begin{eqnarray}
 \tilde m^2_Q(t) &=& \tilde m^2_Q(0)
-\frac{8}{9} ( 1 - Z^2_3 ) \tilde m^2_g(0) - \frac{1}{3}
                  \Delta m^2_t
  \\
 \tilde m^2_T(t) &=& \tilde m^2_T(0)
-\frac{8}{9} ( 1 - Z^2_3 ) \tilde m^2_g(0) - \frac{2}{3}
                  \Delta m^2_t
  \\
 \tilde m^2_B(t) &=& \tilde m^2_B(0)
-\frac{8}{9} ( 1 - Z^2_3 ) \tilde m^2_g(0)
  \\
 \tilde m^2_{H_2}(t) &=& \tilde m^2_{H_2}(0) - \Delta m^2_t \ .
\end{eqnarray}
We recall that the soft mass terms for the squarks of other generations
($\tilde m_{Q_i}, \tilde m_{T_i}, \tilde m_{B_i}$, with $i=1,2$)
and ($\tilde m_{L_i}, \tilde m_{\nu_i}, \tilde m_{E_i}$, with $i=1,2,3$)
for the sleptons have been considered identical to
$\tilde m_Q, \tilde m_T, \tilde m_B$, for simplicity.

\end{document}